\documentclass[twocolumn,aps,prl,floatfix,showpacs]{revtex4}
\usepackage{graphicx}
\begin{document}
\title{Magnetic mechanism of quasiparticle pairing in hole-doped cuprate
    superconductors}
\author{R.S. Markiewicz and A. Bansil}
\affiliation{Physics Department, Northeastern University, Boston MA 02115}
%
\date{\today}
\begin{abstract}

We have computed $\alpha^2F$'s for the hole-doped cuprates within the 
framework of the one-band Hubbard model, where the full magnetic response 
of the system is treated properly. The d-wave pairing weight $\alpha^2F_d$ 
is found to contain not only a low energy peak due to excitations near 
$(\pi ,\pi )$ expected from neutron scattering data, but to also display 
substantial spectral weight at higher energies due to contributions from 
other parts of the Brillouin zone as well as pairbreaking ferromagnetic 
excitations at low energies. The resulting solutions of the Eliashberg 
equations yield transition temperatures and gaps comparable to the 
experimentally observed values, suggesting that magnetic excitations of 
both high and low energies play an important role in providing the pairing 
glue in the cuprates.

\end{abstract}

\maketitle
%

\section{I. Introduction}

Since the superconducting state in the cuprates evolves from the doping of 
a Mott insulator, it is natural to conjecture that the pairing is driven 
by magnetic fluctuations rather than by phonons. Quantum Monte Carlo (QMC)
calculations provide evidence for d-wave pairing\cite{DCA,DCA2}, where the 
pairing bosons reside predominantly in the (transverse) spin channel. 
Recent debate in this connection has centered on whether or not the 
magnetic resonance peak is strong enough to account for the condensation 
energy\cite{resy1,resn}.  Although recent estimates seem to be 
affirmative\cite{resy2}, they do not take into account competing 
pairbreaking effects which enter the Eliashberg equations\cite{Elia}. Also, 
there are arguments that high-energy excitations play a 
role\cite{PWA,glue,CGP}.  Here 
we report a computation of $\alpha^2F$'s for the hole-doped cuprates based 
on the one-band Hubbard model, where the full magnetic response of the 
system is included, and the Eliashberg equations are then solved 
selfconsistently to obtain the superconducting properties over a wide 
range of dopings and temperatures. The resulting transition temperatures 
and pairing gaps are found to be comparable to experimental values, showing 
clearly the viability of the magnetic mechanism in the cuprates.  We find 
that excitations at both high and low energies are important.

Early calculations of magnetic pairing in the cuprates employed parametrized 
models of the susceptibility. The analysis of Radtke, et al. (RULN)\cite{RUIN} 
invokes neutron scattering measurements, while that of Millis, et al. 
(MMP)\cite{MMP} is based on NMR data. The model $\alpha^2F$'s so obtained 
lead to divergent predictions concerning the feasibility of magnetic 
mechanism\cite{SchNo}. Our d-wave pairing weight $\alpha^2F_d$ contains 
not only a low-energy peak (LEP) from near-$(\pi ,\pi )$ scattering, but 
also an additional high energy feature (HEF) extending to $\sim$1.5 eV 
dominated by other regions of the Brillouin zone (BZ), as well as a 
significant pairbreaking contribution at low energies from ferromagnetic 
fluctuations. The HEF, which was missing in the RULN and MMP models turns 
out to be crucially important in producing high transition temperatures 
and pairing gaps. The pairbreaking terms begin to dominate as the Fermi 
energy approaches the van Hove singularity (VHS) with increasing doping 
and can lead to the loss of superconductivity.

Our study bears on the recently discovered `waterfall' or high-energy kink 
(HEK) features observed over 0.3-1.2 eV range in the angle-resolved 
photoemission (ARPES) spectra of a number of cuprates. The magnetic 
susceptibility underlying our computation of $\alpha^2F$'s yields 
self-energies and dispersions consistent with the waterfall 
effects\cite{water3,MJM}, suggesting that the boson responsible for the 
waterfall effects is also a key player in generating significant pairing 
weight in $\alpha^2F_d$ and high condensation energy in the cuprates.

The calculations are based on a one-band Hubbard Hamiltonian, 
extended to include pairing interaction. Specifically, in terms of 
susceptibility $\chi_0$ and the Hubbard on-site repulsion $U$, we use the 
singlet pairing potential\cite{flex2}
\begin{equation}
V_s={U\over 1-U^2\chi_0^2(p'-p)}+{U^2\chi_0(p'+p)\over 1-U\chi_0(p'+p)}
\label{eq:1}
\end{equation}
and the mass renormalization potential (Eq.~[8] of 
Ref.~\onlinecite{flex2}(a))
\begin{equation}
V_z={U^2\chi_0(p'-p)\over 1-U^2\chi_0^2(p'-p)}
+{U^3\chi_0^2(p'-p)\over 1-U\chi_0(p'-p)},
\label{eq:3} 
\end{equation}
where $p$ and $p'$ are the electron momenta, which are constrained to lie 
on the Fermi surface.  Here $V_z$ [$V_s$] is the potential contributing to the normal 
[anomalous] part of the self energy.  These expressions have been found to give 
transition temperatures in good agreement with QMC results\cite{DCA2}.  The resulting 
coupling constants in various pairing channels $\alpha$ are
\begin{equation}
\bar\lambda_{\alpha}=-\int\int d^2pd^2p'\tilde g_{\alpha}(p)\tilde g_{\alpha}(p')
V(p,p',\omega =0)
\label{eq:4}
\end{equation}
where $V=V_s$ for the even parity channels. The normalized weighting 
function
$\tilde g_{\alpha}=g_{\alpha}(p)/(N_0|v_p|)$, where $v_p$ is the Fermi velocity and 
$N_0^2=(2\pi)^3\int{g_{\alpha}(p)^2d^2p/|v_p|}$.  The $g_{\alpha}$ 
are weighting functions of various symmetry\cite{flex2}, of which the most important are 
the lowest harmonics of $s$-wave and $d_{x^2-y^2}$ symmetry, with $g_s=1$ and 
$g_d=cos(p_xa)-cos(p_ya)$.  We also define the coupling constant 
$\lambda_z$ via the $s$-wave version of Eq.~\ref{eq:4} with $V=V_z$.
Then the effective BCS coupling is $\lambda_{\alpha}=\bar\lambda_{\alpha}/
(1+\lambda_z)$.  The symmetrized Eliashberg functions then are
\begin{equation}
\alpha^2F_{\alpha}(\omega)=-{1\over\pi}\int\int d^2pd^2p'\tilde g_{\alpha}(p)\tilde 
g_{\alpha}(p')V''(p,p',\omega),
\label{eq:5}
\end{equation}
where $V''$ is the imaginary part of the corresponding $V$.

\begin{figure}
            \resizebox{7.8cm}{!}{\includegraphics{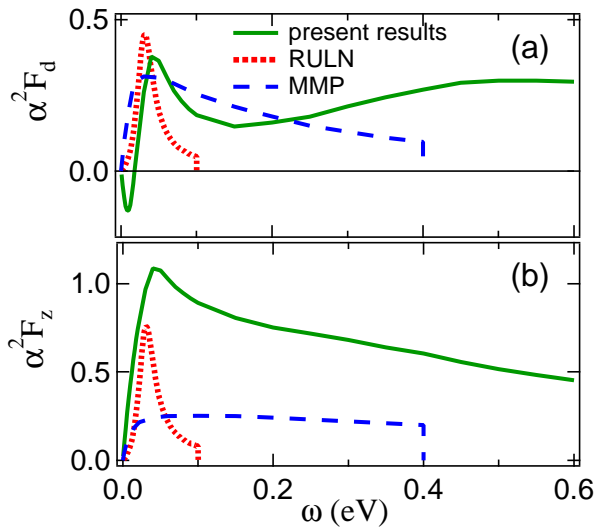}}
\caption{(Color online)
Eliashberg functions $\alpha^2F_d$ and $\alpha^2F_z$ for hole doping 
$x=0.30$ obtained in this work (green line) are compared with
results of Refs.~\protect\onlinecite{RUIN} (red dotted line) and 
~\protect\onlinecite{MMP} (blue dashed line).
}
\label{fig:1}
\end{figure}


In the presence of strong magnetic fluctuations Migdal's theorem is not 
obeyed\cite{Mont,Trem}.  We have developed a relatively simple 
approximation 
scheme\cite{water3,Tanmoy6} which can successfully reproduce the pseudogap and 
waterfall phenomena in the normal state of the cuprates over the full 
doping range.  In the overdoped regime, this scheme reduces to calculating 
the self-energy in GW approximation, using a reduced $U=3.2t$ and 
dispersion renormalized via $Z_0=2$.\cite{footz} These values of $U$ and 
$Z_0$ yield self-energies in reasonable accord with the QMC 
results\cite{Uofx} and explain the recently observed waterfall effects in 
photoemission spectra of the cuprates.\cite{water3,MJM}  We therefore 
expect these parameters to be most appropriate near $x=0.27$, but to gain 
some understanding of how the band structure would affect 
superconductivity in the absence of pseudogap effects, we solved the 
Eliashberg equations over the full doping range $x=0-0.4$, assuming $Z_0$ 
and $U$ to be doping independent.  A more 
satisfactory procedure would be to 
let $U$ increase in the underdoped regime.  However, in the presence of a 
pseudogap, a tensor system of Eliashberg equations needs to be solved, and 
that is beyond the scope of the present calculation. In short, we
proceed thus by solving 
Eqs.~1 and~2 using $U=3.2t$ and $\chi_0$ renormalized by $Z_0$.  In particular, 
we neglect the additional modifications of Migdal's theorem in the 
superconducting state.  Despite this limitation, our results provide 
a benchmark for the Eliashberg formulation in that we do not invoke 
empirical susceptibilities as has been the case in much of the 
existing literature.


Concerning technical details, we use a tight-binding parametrization of 
the dispersion of Bi2212, with the bilayer splitting neglected.\cite{hop} 
$\chi_0$ is first computed within the RPA scheme throughout the BZ for 
frequencies up to 2.88 eV. $\alpha^2F$'s and the $\lambda$'s are then 
computed from Eqs. 1-4. Fermi surface restricted Eliashberg 
equations\cite{RUIN} are finally used to selfconsistently obtain the gap 
$\Delta (\omega )$ and renormalization $Z(\omega )$ functions, with 
$Z(0)\equiv Z=1+\lambda_z$.\cite{foot2}


\section{II. Pure d-wave solution}

Figure~\ref{fig:1}, which compares our typical d-wave pairing weights 
$\alpha^2F_d$ and $\alpha^2F_z$ with RULN and MMP models, highlights our 
key finding. Our $\alpha^2F_d$ (green line) in (a) displays two clear 
features\cite{foot1}: A low energy peak (LEP) around 40 meV and a broad 
high energy hump-like feature (HEF) extending from $\sim 0.5-1.0$ eV (see 
also Fig. 2(a) below). The LEP arises mainly from the magnetic response 
near $(\pi ,\pi )$, but the HEF is connected with the response from other 
parts of the BZ, particularly near $(\pi ,0)$ and $(\pi /2,\pi /2)$. Our 
LEP in (a) is similar to the weights assumed by RULN and MMP.  This 
resemblance is not surprising since the RULN model\cite{RUIN} was designed 
to match neutron scattering data near $(\pi ,\pi )$, while the NMR data 
utilized by MMP\cite{MMP} is also most sensitive to weight in this part of 
the BZ. It has long been known that neutron scattering near $(\pi ,\pi )$ 
accounts for only about 1/8th of the integrated spectral weight expected 
from a total scattering sum rule\cite{LSC}.  By basing their estimate solely 
on the neutron scattering data near $(\pi ,\pi )$, RULN severely 
underestimated the total d-wave 
pairing weight.  The MMP analysis, based on NMR data, appears to have 
captured more of the weight$-$although still missing the HEF and thus 
underestimating the total weight.  Note also from Fig.~1(b) that both 
models strongly underestimate the renormalization weight $\alpha^2F_z$, 
which opposes the tendency for pairing.

The negative dip in Fig.~1(a) at energies below 20 meV deserves comment. 
This dip reflects pair-breaking magnetic scattering (PBS) near $\Gamma$ 
and was overlooked in the phenomenological RULN and MMP models. For 
simplicity, we will refer to these fluctuations as being ferromagnetic 
(FM), although this is strictly so only at $\Gamma$.  This PBS is related 
to earlier indications of FM instabilities near a VHS\cite{SHluG,AAS}.  A 
similar scenario of competing d-wave pairing vs pairbreaking effects has 
been discussed in the context of electron-phonon pairing\cite{BuSca}.

\begin{figure}
            \resizebox{7.8cm}{!}{\includegraphics{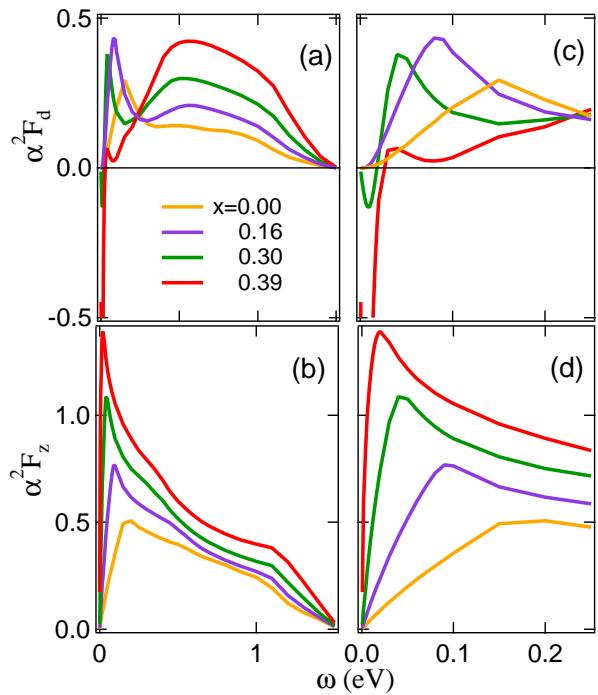}}
\caption{(Color online)
Eliashberg functions $\alpha^2F_d$ and $\alpha^2F_z$ over the doping range 
$x=0.0-0.4$. Lines of various colors refer to different dopings 
(see legend in (a)). Left hand panels (a) and (b) give results over an 
extended frequency range of $0-1.5$ eV, while right hand panels (c) and 
(d) highlight 
the low energy region of $0-250$ meV on an expanded energy scale.
}
\label{fig:2}
\end{figure}

Figure~2 shows how $\alpha^2F$'s evolve with doping. In (a), the pairing 
weight in the high energy feature of $\alpha^2F_d$ is seen to increase 
monotonically with increasing doping, displaying an approximate isosbestic 
point at $\omega\sim 0.24$~eV. In the low energy region in (c), position 
of the peak in $\alpha^2F_d$ shifts to lower energies with increasing 
doping, and the negative pair breaking peak grows dramatically, consistent 
with the suggestion of Kopp et al.\cite{AAS}. The nature of $\alpha^2F_d$ 
is seen to change quite substantially as the Fermi energy approaches the 
VHS at around $x=0.39$. Interestingly, by comparing (c) and (d), the low 
energy peak in $\alpha^2F_z$ is seen to follow that in $\alpha^2F_d$ to 
lower energies with doping.

\begin{figure}
            \resizebox{7.8cm}{!}{\includegraphics{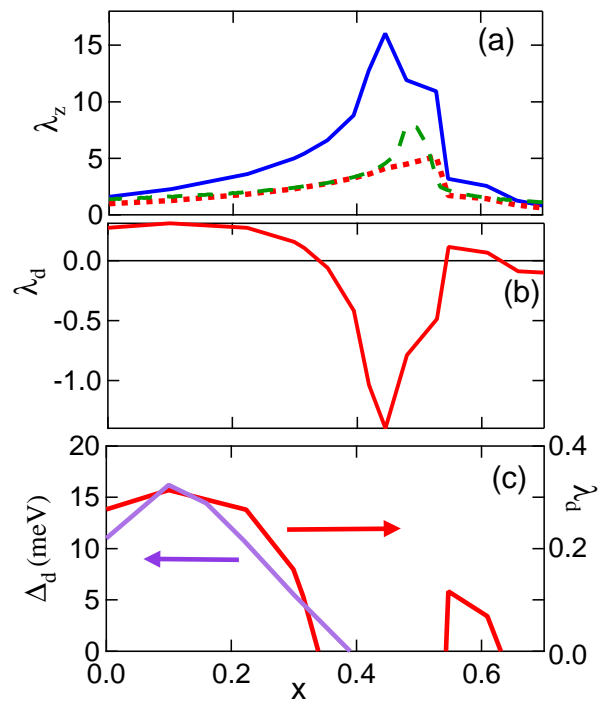}}
\caption{(Color online)
Doping dependence of: (a) $\lambda_z$; (b) $\lambda_d$; and (c) 
$\Delta_d(T=0)$ (left scale) compared to $\lambda_d$ (right scale). In (a) three 
different computations of $\lambda_z$ are compared based on the full $V_z$ of 
Eq.~2 (blue solid line), a simplified $V_{z0}=U^2\chi_0$ (red dotted line), 
and the estimate $N(0)U$ (green dashed line), where $N(0)$ is 
the density-of-states at the Fermi energy.
}
\label{fig:3}
\end{figure}

Figure~\ref{fig:3} shows the doping dependence of $\lambda_z$, $\lambda_d$ 
and the low temperature gap $\Delta_d(T=0)$. Three different estimates of 
$\lambda_z$ are compared in (a) for illustrative purposes. Values based on 
using the bare susceptibility, $V_{z0}=U^2\chi_0$ (red dashed line), are seen to be 
quite similar to the simple estimate $N(0)U$ (green dotted line), where 
$N(0)$ is the density-of-states at the Fermi energy. The full  $V_z$ (blue line) on 
the other hand yields a significant enhancement of $\lambda_z$ over that obtained from 
$\chi_0$, especially near the region of the VHS peak, indicating that the system is 
close to a magnetic instability. Note that $\lambda_d$ is positive for 
dopings less than $\approx$ 0.4, but as the Fermi energy enters the region of 
the VHS with increasing doping, $\lambda_d$ rapidly becomes large and 
negative due to FM fluctuations. (c) shows that this doping 
dependence of $\lambda_d$ is well correlated with that of the pairing gap.
We stress that these results hold for a {\it pure $d_{x^2-y^2}$ order 
parameter.}  Harmonic content plays an important role, as will be discussed 
below, Section III.

\begin{figure}
            \resizebox{7.8cm}{!}{\includegraphics{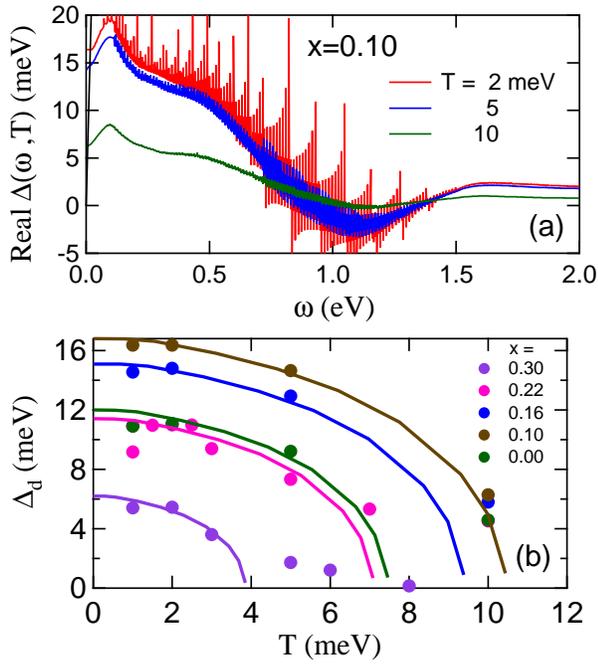}}
\caption{(Color online)
(a) Real part of the gap function $\Delta (\omega ,T)$ at doping $x$=0.10 as a 
function of frequency for a series of temperatures T (see legend). Thin 
black line is the plot of $\Delta=\omega$ used to obtain the low 
energy gap as discussed in the text. 
(b) Computed temperature dependence of the low-energy gap $\Delta_d (T)$ at 
various dopings $x$ (see legend).  
}
\label{fig:4}
\end{figure}

We turn now to discuss our solutions of the Eliashberg equations. 
Following common practice, we proceeded by discretizing the $\alpha^2F$'s 
on the real frequency axis.\cite{KMM} We find that our results are 
sensitive to the number $N_m$ of points in the mesh. For the present 
calculations, based on a 768-point non-uniform mesh over $0-2.88$ 
eV, the gap $\Delta (\omega )$ is approximately converged in the 
low-$\omega$ regime, allowing us to extract $\Delta_d(T)$.  Figure~4(a) 
shows typical results for the real part of $\Delta (\omega )$ for a range 
of temperatures at $x=0.10$. The prominent oscillations in $\Delta 
(\omega )$ curves are the well-known consequence of discretizing 
$\alpha^2F$'s in solving the Eliashberg equations.\cite{MSC}  We define 
the gap by taking the intersection of the $\Delta(\omega)=\omega$ 
line (thin black line in Fig. 4(a)) with the $\Delta (\omega)$ curve. 

Fig.~4(b) shows how the computed low-energy gap $\Delta_d$ evolves with 
temperature at various dopings.  Due to the difficulty of finding 
well-converged solutions when $\Delta$ is small, we calculate $\Delta_d 
(T)$ at a few low temperatures, and use a fit to a $d$-wave BCS gap to 
estimate $T_c$.  We find $2\Delta_d(0)/k_BT_c\sim 3.2$ 
for different dopings. The resulting T$_c$'s are somewhat smaller than QMC 
values\cite{DCA2}, perhaps the effect of a finite $t'$.

\begin{figure}
            \resizebox{6.8cm}{!}{\includegraphics{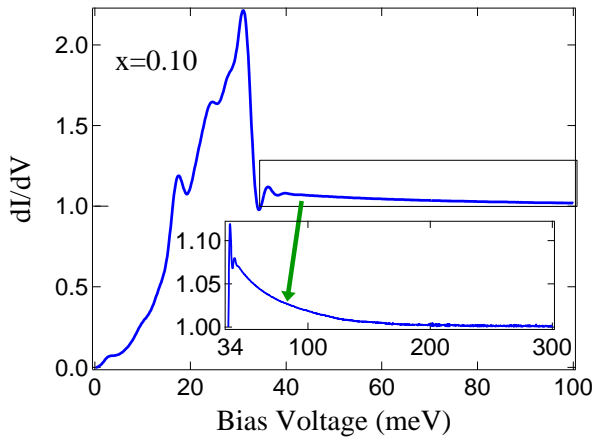}}
\caption{(Color online)
Typical computed SIS tunneling spectrum at $x=0.10$. Inset shows the high 
energy tail on an expanded scale. 
}
\label{fig:5}
\end{figure}

It is striking that the gap features in Fig. 4(a) extend to very high 
energies, raising the obvious question as to how this high-energy tail 
would show up in tunneling spectra.\cite{tunn1} Insight in this regard is 
provided by Fig.~5, where we show a typical tunneling spectrum 
computed\cite{tunn2} within our model. [Tunneling spectra computed at 
other dopings are similar, except that the features scale with 
$\Delta_d$.] The weight in Fig.~5 at energies above the peak-dip-hump 
feature is seen to be quite small with weak energy dependence (see 
inset) and would not be readily observable in the presence of an 
experimental background.

\section{III. Low vs High Energy Pairing Glue}

Within the present model, the LEP and HEF {\it both} play an essential role 
in generating large gaps. For example, at $x=0.3$, the HEF by itself produces 
a gap of only $\sim 0.4$~meV, while the LEP is virtually nonsuperconducting, 
even though the full $\alpha^2F_d$ yields a gap of 5.5 meV.  [To be definite, 
we separate $\alpha^2F$ into LEP and HEF at the minimum in $\alpha^2F$, 
$\omega_{min}=0.3$~eV.]  Similarly, for $x$=0.1, LEP [HEF] by itself has a  
$\sim$3 [0.4] meV gap, with a combined gap of $\sim$17~meV, with 
$\omega_{min}=0.16$~eV.  

This behavior can be readily understood from a 2-$\lambda$ model.\cite{Carb1}
Since this is a purely electronic mechanism, we use a modified Allen-Dynes 
formula\cite{AlD,Carb2}
\begin{equation}
T_c={\omega_{ln}\over 
1.2}exp({-1.04(1+\lambda_z)\over\bar\lambda_d})
\nonumber \\
={\omega_{ln}\over 
1.2}exp({-1.04\over\lambda_d}),
\label{eq:1AD}
\end{equation}
$\Delta (0)=3.54T_c$, with
\begin{equation}
\bar\lambda_d=2\int_0^{\infty}{\alpha^2F(\omega )\over\omega}
\label{eq:2AD}
\end{equation}
and
\begin{equation}
ln(\omega_{ln})={2\over\bar\lambda_d}\int_0^{\infty}ln(\omega ){\alpha^2F(\omega 
)\over\omega}.
\label{eq:3AD}
\end{equation}
The Allen-Dynes equation has a well-known limitation\cite{Carb2} that it 
predicts a maximum $T_c=\omega_{ln}/1.2$, whereas the Eliashberg equations 
have a solution that grows without limit $\sim\sqrt{\bar\lambda}$ as 
$\bar\lambda\rightarrow\infty$.  We find that this leads to an underestimate 
of $\Delta_{LEP}$, while the model provides good estimates for the remaining 
gaps. For instance, for $x=0.1$, $\lambda_{LEP}=\lambda_{HEF}$=0.15, 
$\omega_{ln,LEP}=$~83~meV, $\omega_{ln,HEF}=$~530~meV, so 
$\Delta_{LEF}=$~0.26~meV, and $\Delta_{HEF}=$~1.4~meV.
When both features are combined, 
$\omega_{ln}=$~200~meV and $\lambda_d=0.3$, 
leading to $\Delta_d=19$~meV, in good agreement with the full calculation.  
While the Allen-Dynes model is highly simplified, it does capture the 
observed trend that both peaks contribute significantly.  Physically, the 
effective $\lambda$ is in the weak coupling regime, $\lambda\sim<<1$, so high 
$T_c$ arises from the large $\omega_{ln}$, and the large boost from combining 
LEP and HEF arises since $e^{-1/2\lambda}>>2e^{-1/\lambda}$.
Clearly, an electron-phonon coupling could play a similar 
role in further enhancing $T_c$.

\section{IV. Competing order parameter symmetries}

\begin{figure}
            \resizebox{7.8cm}{!}{\includegraphics{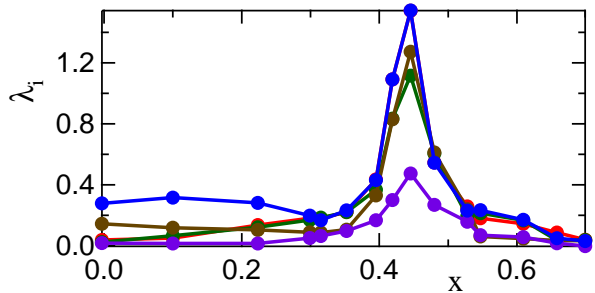}}
\caption{
(Color online)
Doping dependence of: $\lambda_d$ (blue line), $\lambda_s$ (red line), 
$\lambda_{dxy}$ (green line), $\lambda_p$ (violet line), and $\lambda_{sp}$ 
(brown line), calculated from a 15$\times$15 harmonic matrix in each symmetry 
sector.  
}
\label{fig:6}
\end{figure}

The above calculations have been limited to a pure $d$-wave gap symmetry, 
without harmonic content.  In tetragonal symmetry there are five symmetry 
classes of superconducting gap, and each class can involve higher harmonics 
of the given symmetry\cite{Hlub}.  While we have not
solved the tensor Eliashberg equations, it is straightforward to 
generalize 
the $\lambda$ calculations to include harmonic structure and to calculate 
the leading $\lambda$ eigenvalue for each symmetry class. 
The results are shown in Fig.~\ref{fig:6} 
following the analysis of Ref.~\onlinecite{GMV}.  
We see that: (1) The pure-$d$ analysis of Section~II holds in the low 
doping regime; (2) Near the VHS, harmonic content stabilizes d-wave 
symmetry, 
leading to the largest gaps; (3) In this regime, other symmetries 
can become comparable to d-wave.  In particular, there is a tendency 
toward $s$-wave pairing in the overdoped case.

\section{V. Conclusions}

In summary, we have shown that within the present model the d-wave pairing 
weights 
$\alpha^2F_d$ and $\alpha^2F_z$ extend to very high energies of $\sim$ 1 
eV when the magnetic response of the system is properly taken into 
account. The associated superconducting gap is quite substantial, being 
around 16 meV at low dopings.  $\alpha^2F_d$ is found to contain not only the 
expected low energy peak (LEP) below 200 meV, but a previously unrecognized 
high energy feature (HEF) over 0.3-1.2 eV. We find that the LEP 
and HEF both play an important role in yielding a large gap in our 
model.  This suggests that 
electron-phonon coupling could be important for further enhancing $T_c$ as 
suggested by the isotope effect\cite{isot}.  The gap spectrum $\Delta (\omega 
)$ generally extends to the limit of $\alpha^2F$ ($\sim$1.5~eV), and may 
provide insight into a number of anomalous features connected with optical 
properties of the cuprates, summarized for example in Ref.~\onlinecite{Marel}.

The scarcity of high-$T_c$ superconductors arises in part from the fact 
that when superconducting pairing is sufficiently strong, corresponding 
and stronger instabilities arise in other channels.  We have for the most 
part neglected the effects of competing phases, but it is clear that they 
will be significant, both in the underdoped regime and near the VHS.  
Near the VHS pairbreaking ferromagnetic scattering increases sharply, 
strongly suppressing a pure $d$-wave gap.  While we find that inclusion of 
harmonic content could stabilize a d-wave superconductor even at the VHS, 
we have not accounted for a competing FM instability.  Indeed, Storey, et 
al.\cite{TallStorey} find that in Bi-2212 the VHS induces strong pair 
breaking, suppressing superconductivity, so that optimum $T_c$ falls at a 
doping below the VHS.  This is consistent with the evidence for strong FM 
pairbreaking adduced by Kopp et al.\cite{AAS} 

To conclude, we have demonstrated that, when realistic $\alpha^2F$'s are 
used to solve the Eliashberg equations, the magnetic mechanism is capable 
of producing transition temperatures and pairing gaps comparable in size 
to those observed experimentally in the cuprates. The low values of these 
key superconducting properties found in earlier calculations are directly 
attributable to the fact that neutron scattering sees only a fraction of 
the total magnetic spectral weight in these materials.

It is a pleasure to acknowledge useful discussions with Mark Jarrell. This 
work is supported by the US Department of Energy, Basic Energy Sciences, 
Division of Material Science and Engineering contract DE-FG02-07ER46352 
and benefited from the allocation of supercomputer time at NERSC and 
Northeastern University's Advanced Scientific Computation Center (ASCC).



\begin{thebibliography}{99}
\bibitem{hop}The hopping parameters are taken from Ref. 
\onlinecite{water3}: ($t,t',t'',t'''$) = 
(360,--100,35,10)~meV.
\bibitem{foot2}This BCS renormalization factor is the inverse of that 
normally used in Green function theory.
\bibitem{footz}We use $Z_0$ to distinguish the renormalization factor used 
in calculating $\chi_0$; it is not identical to the doping dependent $Z$ 
found from the Eliashberg equations.
\bibitem{foot1}Calculated $\alpha^2F$'s are mainly odd in $\omega$; a 
small even component was subtracted out in the displayed results.
\end{thebibliography}
\end{document}